\documentstyle[11pt,newpasp_jap,twoside]{article}
\reversemarginpar

{

\input epsf
\let\sec=\section
\let\ssec=\subsection

\def\japitem#1{\smallskip\noindent\rlap{#1}\hglue\parindent\hangindent\parindent}
\def\enditem{\smallskip\noindent}

\def\japref{\parskip=0pt\par\noindent\hangindent\parindent
    \parskip =2ex plus .5ex minus .1ex}

\def\gs{\mathrel{\lower0.6ex\hbox{$\buildrel {\textstyle >}
 \over {\scriptstyle \sim}$}}}
\def\ls{\mathrel{\lower0.6ex\hbox{$\buildrel {\textstyle <}
 \over {\scriptstyle \sim}$}}}
\newcount\japequationnum
\global\japequationnum=0
\def\bookdisp#1$${\leftline{\hfill{$\displaystyle#1$}
    \global\advance\japequationnum by 1
    \hfill (\the\japequationnum )}$$}
\everydisplay{\bookdisp}
\def\japsub{\rm\scriptscriptstyle}

\def\kms{{\,\rm km\,s^{-1}}}

\def\hompc{{\,h\,\rm Mpc^{-1}}}
\def\mpcoh{{\,h^{-1}\,\rm Mpc}}

\def\japsub{\rm\scriptscriptstyle}

\def\bk{{\bf k}}
\def\bx{{\bf x}}

\newcommand{\plotter}[2]{\centering \leavevmode \epsfxsize=#2\textwidth \epsfbox{#1}\medskip}
\newcommand{\japplottwo}[2]{\centering \leavevmode 
\epsfxsize=0.49\textwidth \epsfbox{#1}
\hglue 1em
\epsfxsize=0.49\textwidth \epsfbox{#2}
\medskip}

\def\m@th{\mathsurround=0pt }
\def\eqalign#1{\null\,\vcenter{\openup1\jot \m@th
 \ialign{\strut\hfil$\displaystyle{##}$&$\displaystyle{{}##}$\hfil
 \crcr#1\crcr}}\,}

\def\topinsert{\begin{figure}[ht]}

\begin{document}

\title{Studying large-scale structure with the 2dF Galaxy Redshift Survey}
\author{J.A. Peacock$^*$}
\affil{Institute for Astronomy, University of Edinburgh,\\
Royal Observatory, Edinburgh EH9 3HJ, UK
}

\markboth{J.A. Peacock}{2dF Galaxy Redshift Survey}

\begin{abstract}
The 2dF Galaxy Redshift Survey is the first to observe more
than 100,000 redshifts. This allows precise measurements of
many of the key statistics of galaxy clustering,
in particular redshift-space distortions and the large-scale
power spectrum. 
This paper presents the current 2dFGRS results in these areas.
Redshift-space distortions are detected with a high degree of
significance, confirming the detailed Kaiser distortion from
large-scale infall velocities, and measuring the distortion parameter
$\beta \equiv \Omega_m^{0.6}/b = 0.43 \pm 0.07$.
The power spectrum is measured to $\ls 10\%$ accuracy for
$k>0.02 \hompc$, and is well fitted by a CDM model with
$\Omega_m h =0.20 \pm 0.03$ and a baryon fraction of $0.15\pm 0.07$.
A joint analysis with CMB data requires $\Omega_m =0.29 \pm 0.05$,
assuming scalar fluctuations, but no priors on other parameters.
Two methods are used to determine the large-scale bias parameter:
an internal bispectrum analysis yields $b=1.04\pm 0.11$, in very
good agreement with the $b=1.10\pm 0.08$ obtained from a joint
2dFGRS+CMB analysis, again assuming scalar fluctuations. These figures
refer to galaxies of approximate luminosity $2L^*$; luminosity
dependence of clustering is detected at high significance, and is
well described by $b/b^* = 0.85 + 0.15(L/L^*)$.

\end{abstract}

\renewcommand{\thefootnote}{\fnsymbol{footnote}}
\footnotetext[1]{{\sl On behalf of the 2dF Galaxy Redshift Survey team:} Matthew Colless (ANU), 
Ivan Baldry (JHU), Carlton  Baugh (Durham), 
Joss Bland-Hawthorn (AAO), Terry Bridges (AAO),
Russell Cannon (AAO), Shaun Cole (Durham), 
Chris Collins (LJMU), Warrick Couch (UNSW), 
Gavin Dalton (Oxford), Roberto De Propris (UNSW), 
Simon Driver (St Andrews), George Efstathiou (IoA), 
Richard  Ellis (Caltech), Carlos Frenk (Durham), 
Karl Glazebrook (JHU), Carole Jackson (ANU), Ofer Lahav (IoA), Ian Lewis (AAO), 
Stuart Lumsden (Leeds), Steve Maddox (Nottingham), Darren Madgwick (IoA), 
Peder Norberg (Durham), Will Percival (ROE), 
Bruce Peterson (ANU), Will Sutherland (ROE), Keith Taylor (Caltech).}
\renewcommand{\thefootnote}{\arabic{footnote}}

\sec{Aims and design of the 2dFGRS}

The large-scale structure in the galaxy distribution is widely
seen as one of the most important relics from an early stage of
evolution of the universe.
The 2dF Galaxy Redshift Survey (2dFGRS) was designed  to build on previous
studies of this structure, with the following main aims:

\japitem{(1)}To measure the galaxy power spectrum $P(k)$ on scales up to a few
hundred Mpc, bridging the gap between the scales of nonlinear
structure and measurements from the the cosmic microwave background (CMB).

\japitem{(2)}To measure the redshift-space distortion of the large-scale clustering
that results from the peculiar velocity field produced by the mass
distribution. 

\japitem{(3)}To measure higher-order clustering statistics in order to
understand biased galaxy formation, and to test
whether the galaxy distribution on large scales
is a Gaussian random field.

\enditem
The survey is designed around the 2dF multi-fibre spectrograph on the
Anglo-Australian Telescope, which is capable of observing up to 400
objects simultaneously over a 2~degree diameter field of view. 
For details of
the instrument and its performance 
see {\tt http://www.aao.gov.au/2df/}, and also
Lewis et~al.\ (2002).

The source catalogue for the survey is a revised and extended version of
the APM galaxy catalogue (Maddox et~al.\ 1990a,b,c); this
includes over 5~million galaxies down to $b_{\japsub J}=20.5$ in both
north and south Galactic hemispheres over a region of almost
$10^4\, {\rm deg}^2$ (bounded approximately by declination $\delta \leq+3^\circ$
and Galactic latitude $b\gs 20^\circ$). 
This catalogue is
based on Automated Plate Measuring machine (APM) scans of 390 plates
from the UK Schmidt Telescope (UKST) Southern Sky Survey. The $b_{\japsub J}$
magnitude system for the Southern Sky Survey is defined by the response
of Kodak IIIaJ emulsion in combination with a GG395 filter,
and is related to the Johnson--Cousins system by $b_{\japsub J} = B -0.28(B-V)$.
The photometry of the catalogue is calibrated with numerous
CCD sequences and has a precision of approximately 0.15~mag for galaxies
with $b_{\japsub J}=17$--19.5. The star-galaxy separation is as described in
Maddox et~al.\ (1990b), supplemented by visual validation of each galaxy
image.

\begin{figure}[ht]
\plotter{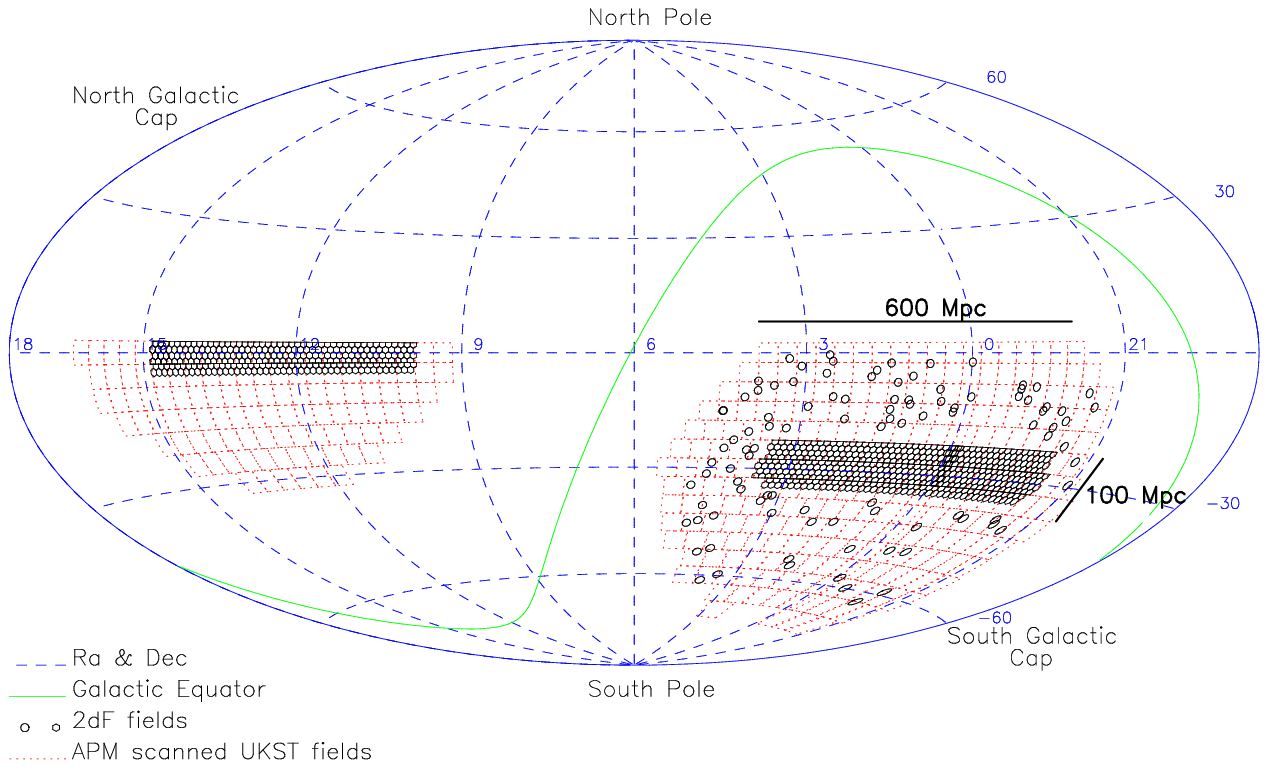}{0.7}
\caption{The 2dFGRS fields (small circles) superimposed on the APM
catalogue area (dotted outlines of Sky Survey plates). There are
approximately 140,000 galaxies in the $75^\circ\times15^\circ$
southern strip centred on the SGP, 70,000 galaxies in the
$75^\circ\times7.5^\circ$ equatorial strip, and 40,000 galaxies in
the 100 randomly-distributed 2dF fields covering the whole area of the
APM catalogue in the south.}
\end{figure}

The survey geometry is shown in Figure~1, and consists of two contiguous
declination strips, plus 100 random 2-degree fields. One strip is in the
southern Galactic hemisphere and covers approximately
75$^\circ$$\times$15$^\circ$ centred close to the SGP at
($\alpha, \delta$)=($01^h$,$-30^\circ$); the other strip is in the northern
Galactic hemisphere and covers $75^\circ \times 7.5^\circ$ centred at
($\alpha, \delta$)=($12.5^h$,$+0^\circ$). The 100 random fields are spread
uniformly over the 7000~deg$^2$ region of the APM catalogue in the
southern Galactic hemisphere. At the median redshift of the survey
($\bar{z}=0.11$), $100\mpcoh$ subtends about 20~degrees, so the two strips
are $375\mpcoh$ long and have widths of $75\mpcoh$ (south) and $37.5\mpcoh$
(north). 

The sample is limited to be brighter than an extinction-corrected
magnitude of $b_{\japsub J}=19.45$ (using the extinction maps of Schlegel et~al.\
1998). This limit gives a good match between the density on the sky of
galaxies and 2dF fibres. Due to clustering, however, the number in a
given field varies considerably. To make efficient use of 2dF, we employ
an adaptive tiling algorithm to cover the survey area with the minimum
number of 2dF fields. With this algorithm we are able to achieve a 93\%
sampling rate with on average fewer than 5\% wasted fibres per field.
Over the whole area of the survey there are in excess of 250,000
galaxies.

\begin{figure}[ht]
\plotter{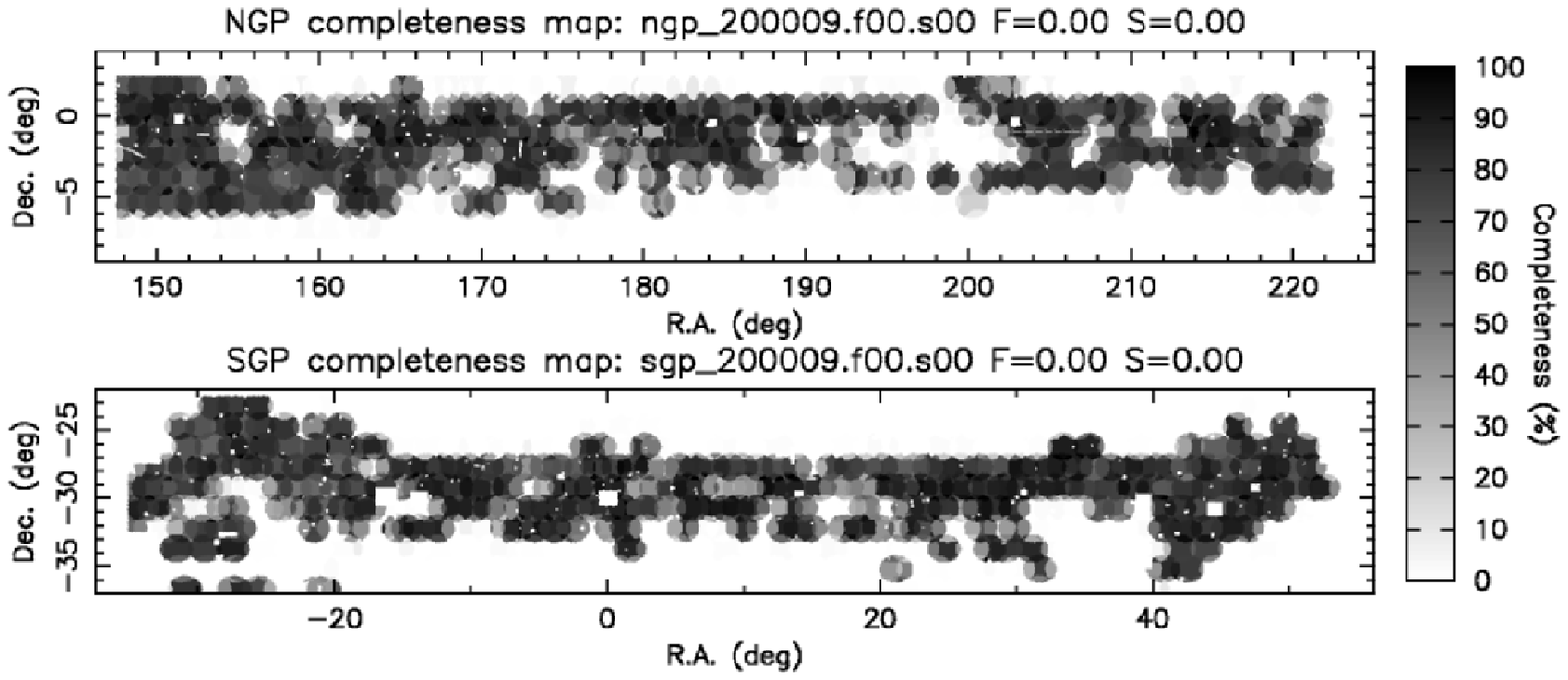}{1.0}
\caption{The completeness as a function of position on the sky, as at January 2000. The
circles are individual 2dF fields (`tiles'). Unobserved tiles result in
low completeness in overlap regions. Rectangular holes are omitted
regions around bright stars. The lack of a complete set of overlapping tiles
means that the completeness relative to the parent survey varies by over a factor of 2. 
However, such variations in the effective mask can be allowed for in statistical analyses.}
\end{figure}

\sec{Survey Status}

By the end of 2001,
observations had been made of 866 fields,
yielding redshifts and identifications for 224,851 galaxies, 13630 stars
and 176 QSOs, at an overall completeness of 93\%. 
The galaxy redshifts are assigned a quality flag from 1 to 5,
where the probability of error is highest at low $Q$. Most analyses
are restricted to $Q\ge 3$ galaxies, of which there are currently
213,703.
Data-taking will continue in 2002, but concentrating
on increasing the completeness of the existing survey zone;
the total number of $Q\ge 3$ galaxies is expected to asymptote
to about 230,000. An interim data release took place in July 2001,
consisting of approximately 100,000 galaxies (see Colless et al. 2001
for details). A public release of the full photometric and spectroscopic
database is scheduled for July 2003.

\begin{figure}[ht]
\plotter{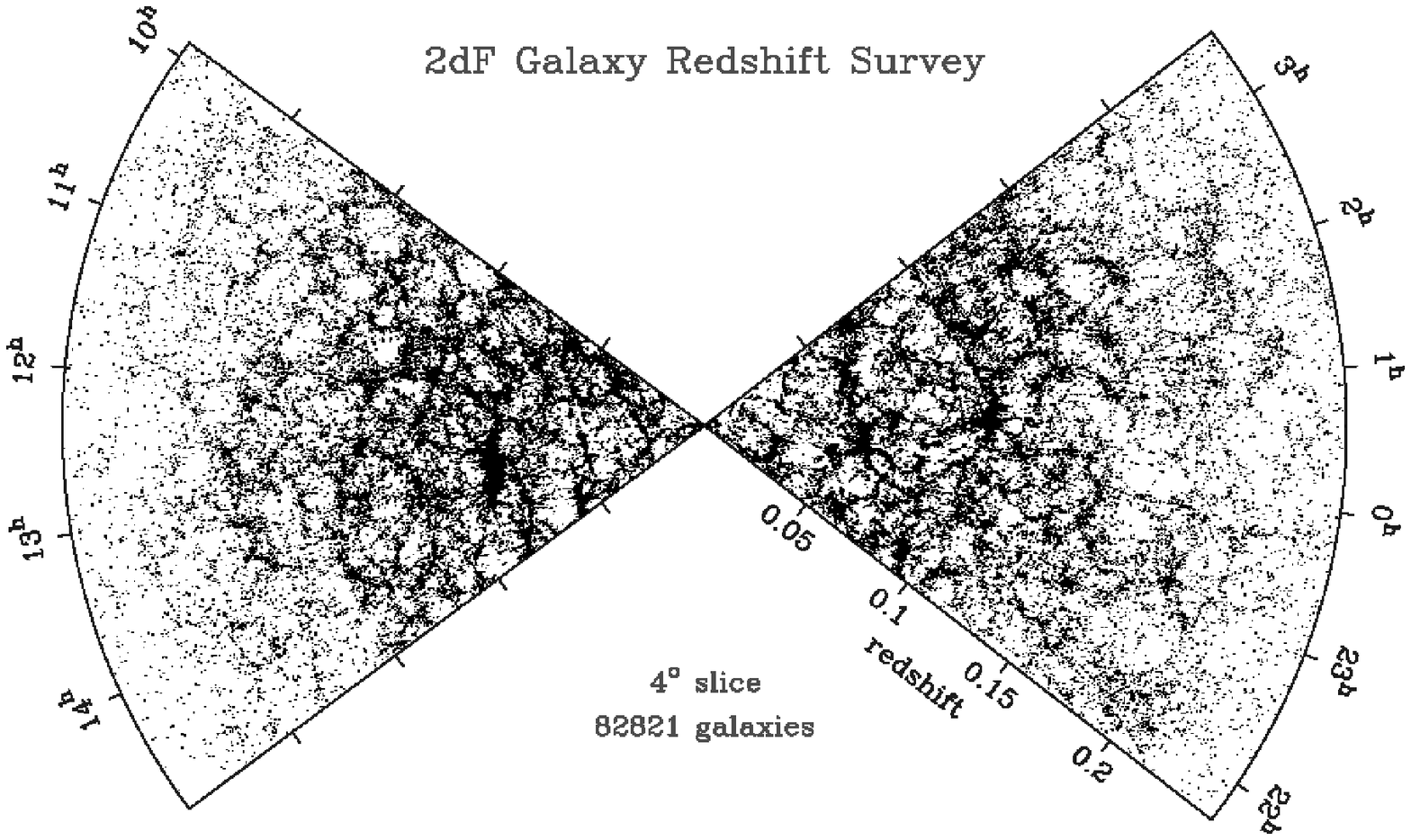}{1.0}
\caption{The distribution of galaxies in part of the 2dFGRS, drawn from 
a total of 213,703 galaxies:
slices $4^\circ$ thick, centred at declination
$-2.5^\circ$ in the NGP and $-27.5^\circ$ in the SGP.
This image reveals a wealth of detail, including
linear supercluster features, often nearly perpendicular
to the line of sight. The interesting question
to settle statistically is whether such transverse features
have been enhanced by infall velocities.}
\end{figure}

The Colless et al. (2001) paper details the practical steps that
are necessary in order to work with a survey of this sort.
The 2dFGRS does not consist of a simple region sampled with 100\%
efficiency, and it is therefore necessary to use a number of 
masks in order to interpret the data. Two of these concern the
input catalogue: the boundaries of this catalogue, including
`drilled' regions around bright stars where galaxies could not
be detected; also, revisions to the photometric calibration
mean that in practice the survey depth varies slightly with position on the sky.
The most important mask, however, arises from the way in which
the sky is tessellated into 2dF tiles.
The adaptive tiling algorithm is efficient, and yields uniform
sampling in the final survey. However, at any intermediate stage,
missing overlaps mean that the sampling fraction has fluctuations,
as illustrated in Figure~2. 
This variable sampling makes quantification of the large scale
structure more difficult, particularly for any analysis requiring relatively
uniform contiguous areas. However, the effective survey `mask' can
be measured precisely enough that it can be allowed for in
low-order analyses of the galaxy distribution. Figure~2 shows the
mask at an earlier stage of the survey, appropriate for some of the
first 2dFGRS  analyses. The final database will have a more uniform
mask, but it will always be necessary to allow for these
sampling nonuniformities.

Despite these issues, the 2dFGRS now yields a strikingly detailed
and complete view of the galaxy distribution over large cosmological volumes.
This is illustrated in
Figure~3, which shows the projection of a subset of
the galaxies in the northern and southern strips onto $(\alpha,z)$
slices. In contemplating this picture, it is worth remembering
the decades of effort that have been invested in establishing
the reality of large-scale structure. Mapping these
structures clearly is without doubt a historically important
intellectual step for the human race, and 
it is a privilege to contribute to this process.

\begin{figure}[ht]
\plotter{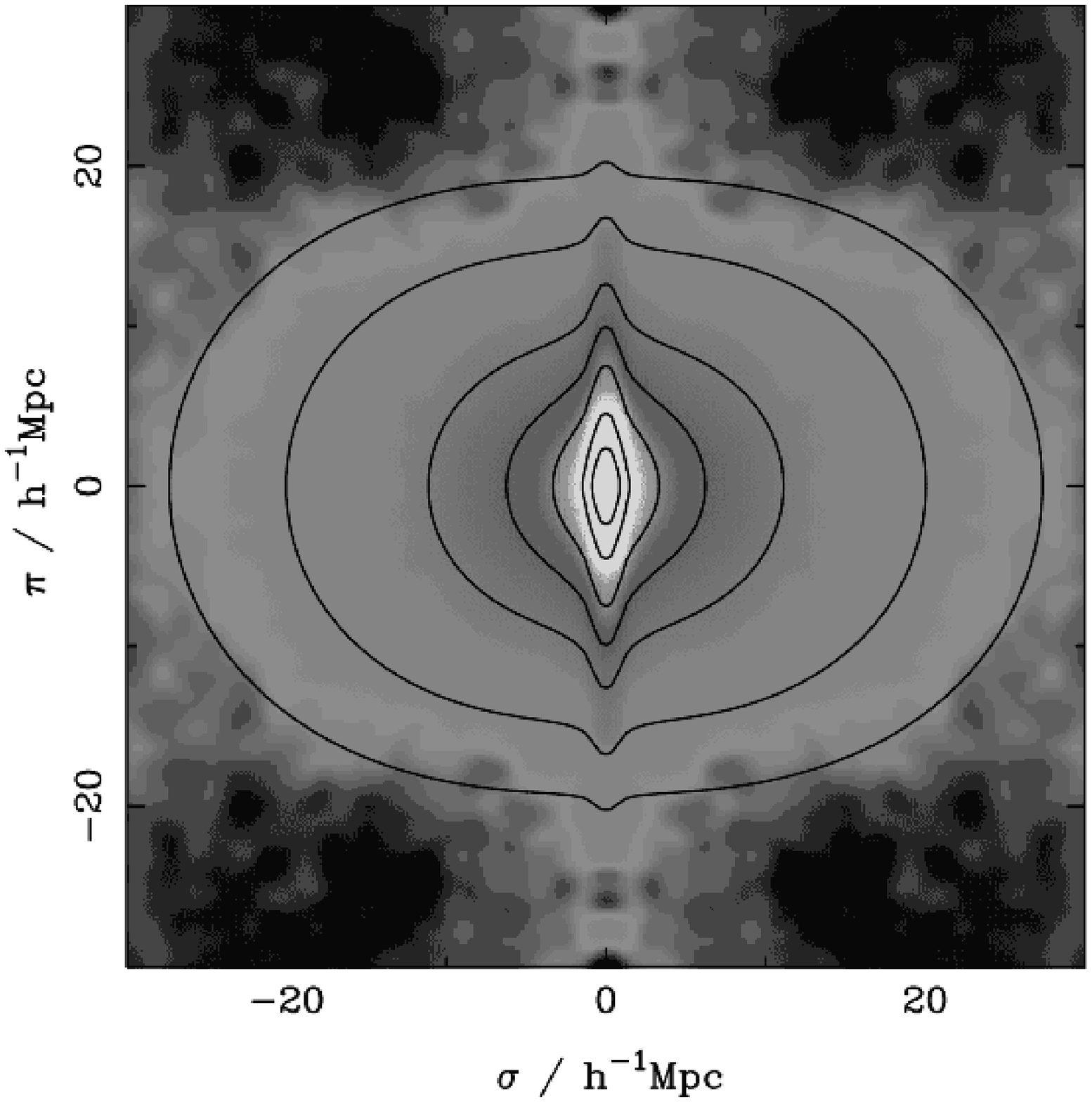}{0.6}
\caption{The galaxy correlation function $\xi(\sigma,\pi)$ as a function
of transverse ($\sigma$) and radial ($\pi$) pair separation is shown as
a greyscale image. It was computed in $0.2\mpcoh$ boxes and then smoothed
with a Gaussian having an rms of $0.5\mpcoh$. The contours are for a model
with $\beta=0.4$ and $\sigma_p=400\kms$, and are plotted at $\xi=10$, 5,
2, 1, 0.5, 0.2 and 0.1.}
\end{figure}

\sec{Redshift-space correlations}

The simplest statistic for studying clustering in the galaxy
distribution is the the two-point correlation function,
$\xi(\sigma,\pi)$. This measures the excess probability over random of
finding a pair of galaxies with a separation in the plane of the sky
$\sigma$ and a line-of-sight separation $\pi$. Because the radial
separation in redshift space includes the peculiar velocity as well as
the spatial separation, $\xi(\sigma,\pi)$ will be anisotropic. On small
scales the correlation function is extended in the radial direction due
to the large peculiar velocities in non-linear structures such as groups
and clusters -- this is the well-known `Finger-of-God' effect. On large
scales it is compressed in the radial direction due to the coherent
infall of galaxies onto mass concentrations -- the Kaiser effect (Kaiser
1987).

To estimate $\xi(\sigma,\pi)$ we compare the observed count of galaxy
pairs with the count estimated from a random distribution following the
same selection function both on the sky and in redshift
as the observed galaxies. We apply optimal weighting to
minimise the uncertainties due to cosmic variance and Poisson noise.
This is close to equal-volume weighting out to an adopted redshift
limit of $z=0.25$. We have tested our results and found them to be
robust against the uncertainties in both the survey mask and the
weighting procedure.
The redshift-space correlation function for the 2dFGRS 
computed in this way is shown in Figure~4.
The correlation-function results display very clearly
two signatures of redshift-space distortions.
The `fingers of God' from small-scale random
velocities are very clear, as indeed has been the case
from the first redshift surveys (e.g. Davis \& Peebles 1983).
However, this is the first time that the 
large-scale flattening from coherent infall has been seen in detail.

\begin{figure}[ht]
\japplottwo{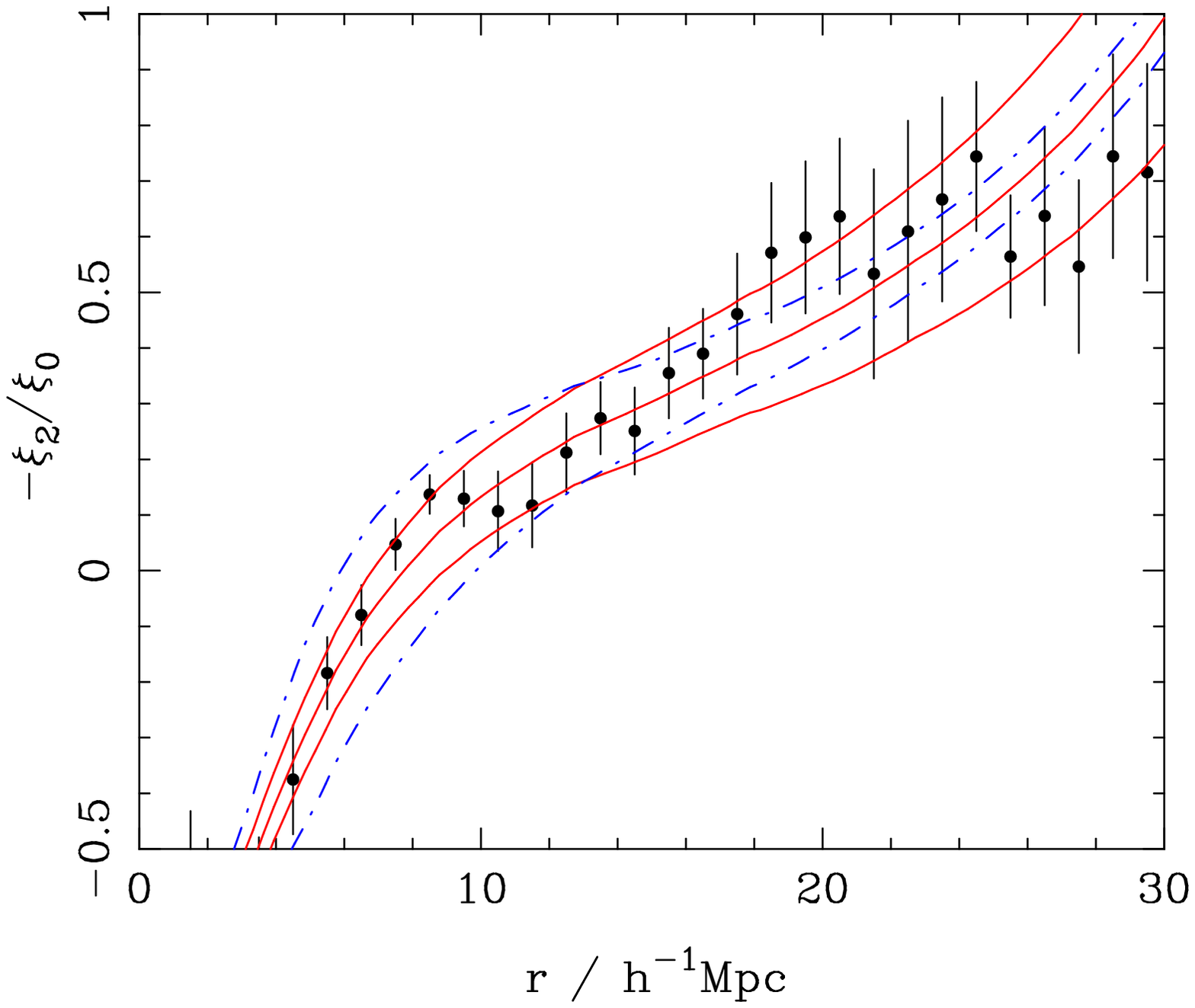}{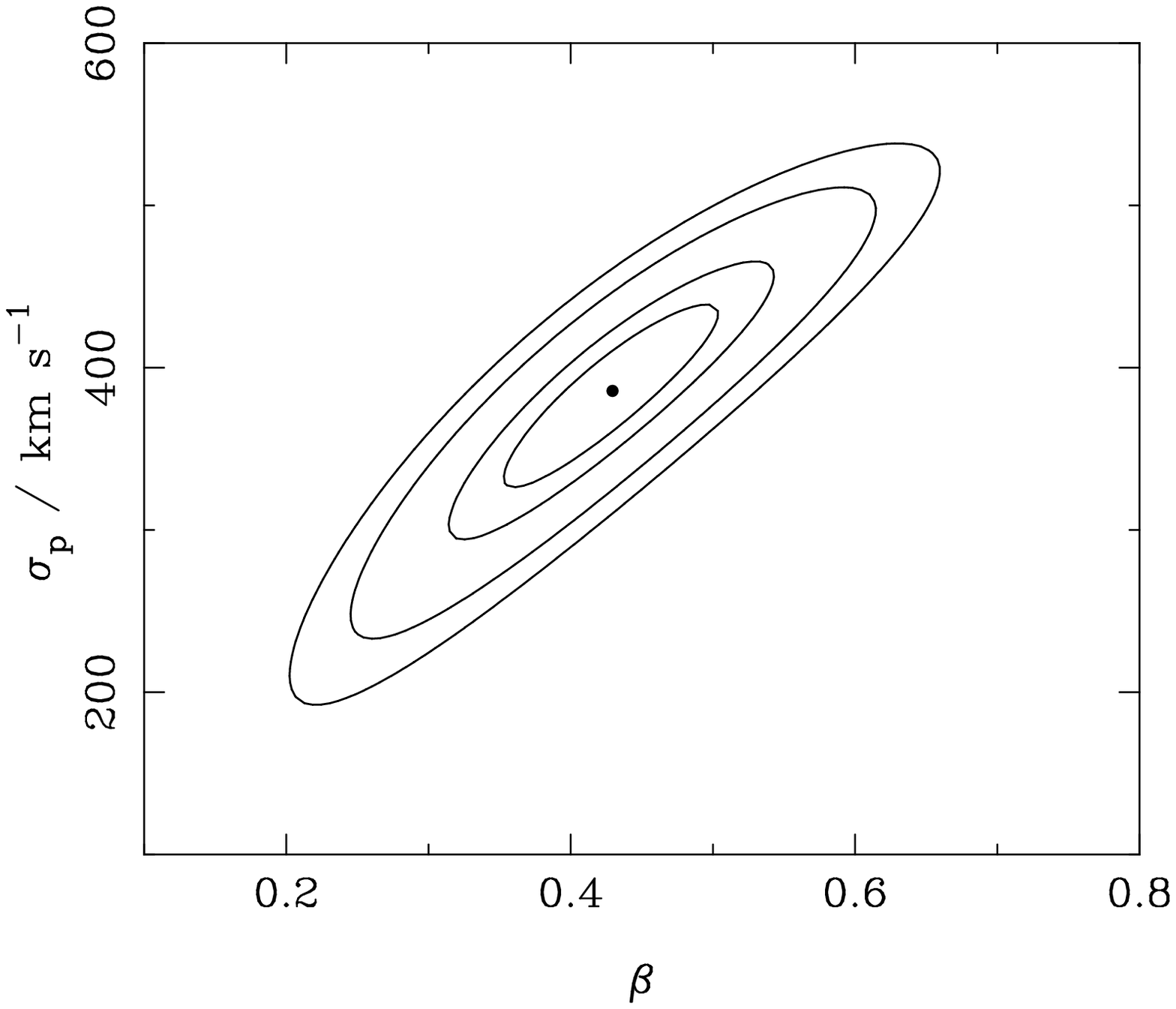}
\caption{(a)~The compression of $\xi(\sigma,\pi)$ as measured by its
quadrupole-to-monopole ratio, plotted as $-\xi_2/\xi_0$. The solid lines
correspond to models with $\sigma_p=400\kms$ and (bottom to top)
$\beta=0.3$,0.4,0.5, while the dot-dash lines correspond to models with
$\beta=0.4$ and (top to bottom) $\sigma_p=300,400,500\kms$.
(b)~Likelihood contours for $\beta$ and $\sigma_p$ from the model fits.
The inner contour is the one-parameter 68\% confidence ellipse; the
outer contours are the two-parameter 68\%, 95\% and 99\% confidence
ellipses. The central dot is the maximum likelihood fit, with
$\beta =0.43$ and $\sigma_p=385\kms$.}
\end{figure}

The degree of large-scale flattening is
determined by the total mass density parameter, $\Omega_m$, and the
biasing of the galaxy distribution.
On large scales, it should be correct to assume a linear bias model,
with correlation functions $\xi_g(r) = b^2 \xi(r)$,
so that the redshift-space distortion on large scales depends on
the combination $\beta \equiv \Omega_m^{0.6}/b$. On these scales, linear
distortions should also be applicable, so we expect to see the
following quadrupole-to-monopole ratio in the correlation function:
$$
\frac{\xi_2}{\xi_0} =
\frac{3+n}{n}\frac{4\beta/3+4\beta^2/7}{1+2\beta/3+\beta^2/5}
$$ 
(e.g. Hamilton 1992),
where $n$ is the power spectrum index of the fluctuations, $\xi \propto
r^{-(3+n)}$. This is modified by the Finger-of-God effect, which is
significant even at large scales and dominant at small scales. 
The effect can be modelled by introducing a parameter
$\sigma_p$, which represents the rms pairwise velocity dispersion of
the galaxies in collapsed structures, $\sigma_p$ (see e.g. Ballinger et al. 1996).
Full details of the fitting procedure are given in Peacock et al. (2001).

Figure~5a shows the variation in $\xi_2/\xi_0$ as a function of scale.
The ratio is positive on small scales where the Finger-of-God effect
dominates, and negative on large scales where the Kaiser effect
dominates. The best-fitting model (considering only the quasi-linear
regime with $8 < r < 25\mpcoh$) has $\beta\simeq 0.4$ and
$\sigma_p \simeq 400\kms$; the likelihood contours are
shown in Figure~5b. Marginalising over $\sigma_p$, the best estimate of
$\beta$ and its 68\% confidence interval is
$$
\beta=0.43\pm0.07
$$
This is the first precise measurement of $\beta$ from redshift-space
distortions; previous studies have shown the effect to exist
(e.g. Hamilton, Tegmark \& Padmanabhan 2000; Taylor et al. 2001; 
Outram, Hoyle \& Shanks 2001), but achieved little more than 3$\sigma$
detections.

Our measurement of $\Omega^{0.6}/b$ would thus imply 
$\Omega=0.36 \pm 0.10$ if $L^*$ galaxies are unbiased, but it is difficult to justify
such an assumption in advance. We discuss below two methods by which the
bias parameter may be inferred, which do in fact favour a
low degree of bias. Nevertheless, there are other uncertainties
in converting a measurement of $\beta$ to a figure for $\Omega$.
The 2dFGRS has a median redshift of 0.11; with weighting,
the mean redshift in the present analysis is $\bar z =0.17$,
and our measurement should be interpreted as $\beta$ at
that epoch. 
The optimal weighting also means that our mean
luminosity is high: it is approximately
1.9 times the characteristic luminosity, $L^*$,  of the overall
galaxy population (Folkes et al. 1999; Madgwick et al. 2001). 
This means that we need to quantify the luminosity dependence
of clustering.

\sec{Real-space clustering and its dependence on luminosity}

The dependence of galaxy clustering on luminosity is an
effect that was controversial for a number of years.
Using the APM-Stromlo redshift survey, Loveday et al. (1995)
claimed that there was no trend of clustering amplitude
with luminosity, except possibly at the very lowest
luminosities. In contradiction, the SSRS study of
Benoist et al. (1996) suggested that the strength of
galaxy clustering increased monotonically with luminosity, with
a particularly marked effect for galaxies above $L^*$.
The latter result was arguably more plausible, based on
what we know of luminosity functions and morphological
segregation. It has been clear for many years that elliptical
galaxies display a higher correlation amplitude than
spirals (Davis \& Geller 1976). Since ellipticals are also
more luminous on average, as shown most clearly by the
2dFGRS luminosity function results (Folkes et al. 1999;
Madgwick et al. 2001), some trend with luminosity is to
be expected, but the challenge is to detect it.

\begin{figure}[ht]
\plotter{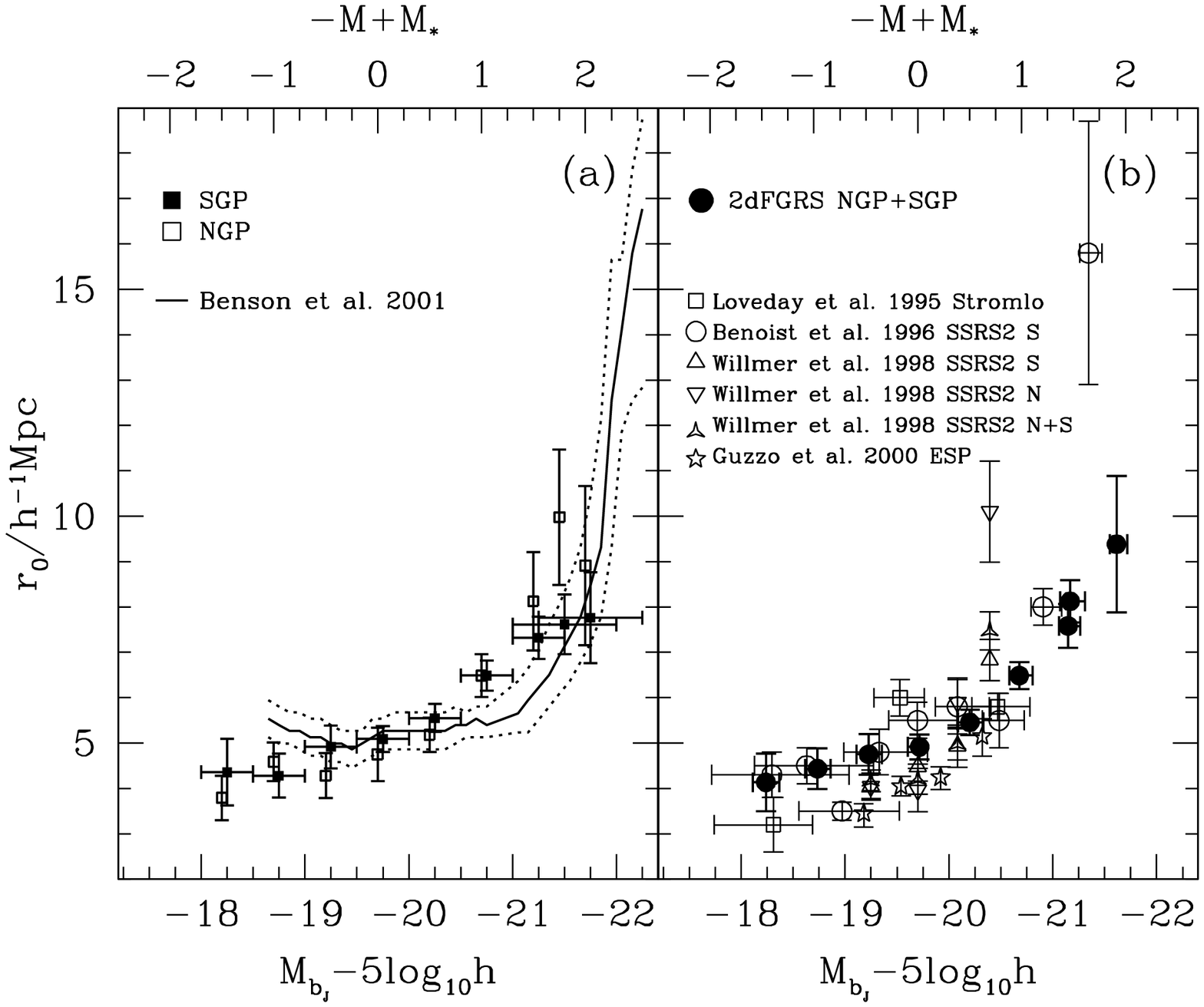}{0.7}
\caption{(a) The correlation length in real space as a function of 
absolute magnitude. 
The solid line shows the predictions of the semi-analytic 
model of Benson et al. (2001), computed in a series of overlapping 
bins, each $0.5$ magnitudes wide. The dotted curves show an 
estimate of the errors on this prediction, including the relevant sample variance
for the survey volume.
(b) The real space correlation length estimated combining 
the NGP and SGP (filled circles). 
The open symbols show a selection of recent data from other studies.
}
\end{figure}

The difficulty with measuring the dependence of $\xi(r)$ on
luminosity is that cosmic variance can mask the signal of
interest. It is therefore important to analyse volume-limited samples
in which galaxies of different luminosities are compared
in the same volume of space. This requires the rejection of many
distant high-$L$ galaxies, and so the strategy is only feasible
with a survey of the size of the 2dFGRS. This comparison was undertaken
by Norberg et al. (2001a), who measured real-space correlation functions
via the projection $\Xi(\sigma) = \int \xi(\sigma,\pi)\; d\pi$,
demonstrating that it was possible to
obtain consistent results in both NGP and SGP.
A very clear detection of luminosity-dependent clustering was achieved,
as shown in Figure~6. The results
can be described by a linear dependence of effective bias
parameter on luminosity:
$$
b/b^* = 0.85 + 0.15\,(L/L^*),
$$
and the scale-length of the real-space correlation function for $L^*$
galaxies is approximately $r_0=4.8 \mpcoh$. There is thus a
significant difference in the clustering amplitude of $L^*$
galaxies and the $L\simeq 1.9 L^*$ characteristic of an
optimally weighted sample, and this must be allowed for.
This trend is in qualitative agreement with the results of
Benoist et al. (1996), but in fact these workers gave a stronger
dependence on luminosity than is indicated by the 2dFGRS.
Finally, with spectral classifications, it is possible to
measure the dependence of clustering both on luminosity and
on spectral  type, to see to what extent morphological
segregation is responsible for this result. Norberg et al. (2001b)
show that, in fact, the principal effect seems to be with
luminosity: $\xi(r)$ increases with $L$ for all spectral types.
This is reasonable from a theoretical point of view, in which
the principal cause of different clustering amplitudes is the
mass of halo that hosts a galaxy 
(e.g. Cole \& Kaiser 1989; Mo \& White 1996; 
Kauffman, Nusser \& Steinmetz 1997).

\sec{The 2dFGRS power spectrum}

Perhaps the key aim of the 2dFGRS was to perform an accurate
measurement of the 3D clustering power spectrum, in order
to improve on the APM result,
which was deduced by deprojection of angular
clustering (Baugh \& Efstathiou 1993, 1994). 
The results of this direct estimation of the 3D power
spectrum are shown in Figure~7.
This power-spectrum estimate uses the FFT-based approach
of Feldman, Kaiser \& Peacock (1994), and needs to be interpreted
with care. Firstly, it is a raw redshift-space estimate, so
that the power beyond $k\simeq 0.2 \hompc$ is severely damped
by fingers of God. On large scales, the power is enhanced, both
by the Kaiser effect and by the luminosity-dependent clustering
discussed above. Finally, the FKP estimator yields the
true power convolved with the window function. This
modifies the power significantly on large scales (roughly
a 20\% correction). We have made an approximate correction for
this in Figure~7 by multiplying by the correction factor appropriate
for a $\Gamma=0.25$ CDM spectrum. The precision of the
power measurement appears to be encouragingly high, and the
systematic corrections from the window are well specified.

\begin{figure}[ht]
\plotter{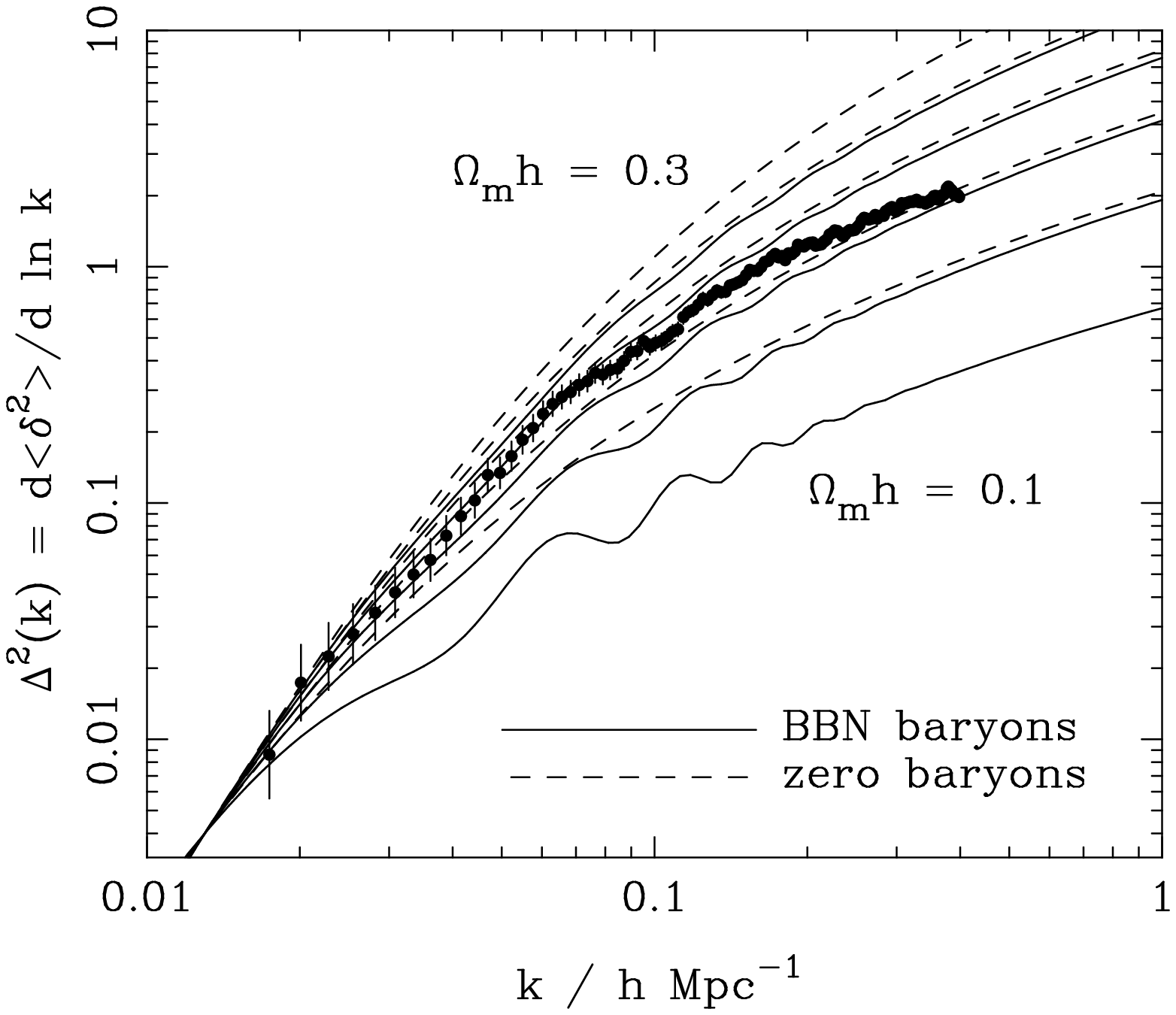}{0.6}
\caption{The 2dFGRS redshift-space dimensionless power spectrum, 
$\Delta^2(k)$,
estimated according to the FKP procedure. The solid points
with error bars show the power estimate. The window
function correlates the results at different $k$ values,
and also distorts the large-scale shape of the power spectrum
An approximate correction for the latter effect has been applied.
The solid and dashed lines show various CDM models, all assuming
$n=1$. For the case with non-negligible baryon content,
a big-bang nucleosynthesis value of $\Omega_b h^2=0.02$ is
assumed, together with $h=0.7$. A good fit is clearly obtained
for $\Omega_m h \simeq 0.2$. Note that the observed power at
large $k$ will be boosted by nonlinear effects, but damped by 
small-scale random peculiar velocities. It appears that these
two effects very nearly cancel, but model fitting is generally
performed only at $k<0.15 \hompc$ in order to avoid these complications.}
\end{figure}

The next task is to perform a detailed fit of physical
power spectra, taking full account of the window effects.
The hope is that we will obtain not only a more precise measurement
of the overall spectral shape, as parameterized by
$\Gamma$, but will be able to move towards more detailed
questions such as the existence of baryonic features in the
matter spectrum (e.g. Meiksin, White \& Peacock  1999).
We summarize here results from the first attempt at this analysis
(Percival et al. 2001).

In order to compare the 2dFGRS power spectrum to members of the
CDM family of theoretical models, it is essential to have a
proper understanding of the full covariance matrix of the data:
the convolving effect of the window function 
causes the power at adjacent $k$ values to be correlated.
This covariance matrix was estimated by applying the survey window to a
library of Gaussian realisations of linear density fields
for a $\Omega_mh=0.2$, $\Omega_b/\Omega_m=0.15$ CDM power spectrum, for
which $\chi^2_{\rm min}=34.4$, given an expected value of $28$. The
best fit power spectrum parameters are only weakly dependent on this
choice. Similar results were obtained using a covariance matrix
estimated from a set of mock catalogues derived from $N$-body
simulations.

It is now possible to explore the space of CDM models,
evaluating likelihood values assuming a Gaussian form for
the likelihood.
The likelihood contours in $\Omega_b/\Omega_m$ versus $\Omega_mh$ for
this fit are shown in Figure~8. At each point in this
surface we have marginalized by integrating the likelihood surface
over the two free parameters, $h$ and the power spectrum
amplitude. The result is not significantly altered if instead, the
modal, or maximum mikelihood points in the plane corresponding to
power spectrum amplitude and $h$ were chosen. The likelihood function
is also dependent on the covariance matrix (which should be allowed to
vary with cosmology), although the consistency of result from
covariance matrices calculated for different cosmologies
shows that this dependence is negligibly small.
Assuming a uniform prior for $h$ over a factor of 2 is arguably
over-cautious, and we have therefore added a Gaussian prior $h=0.7\pm
10\%$. This corresponds to multiplying by
the likelihood from external constraints such as the HST key project
(Freedman et al. 2001); this has only a minor effect on the results.

\begin{figure}[ht]
\plotter{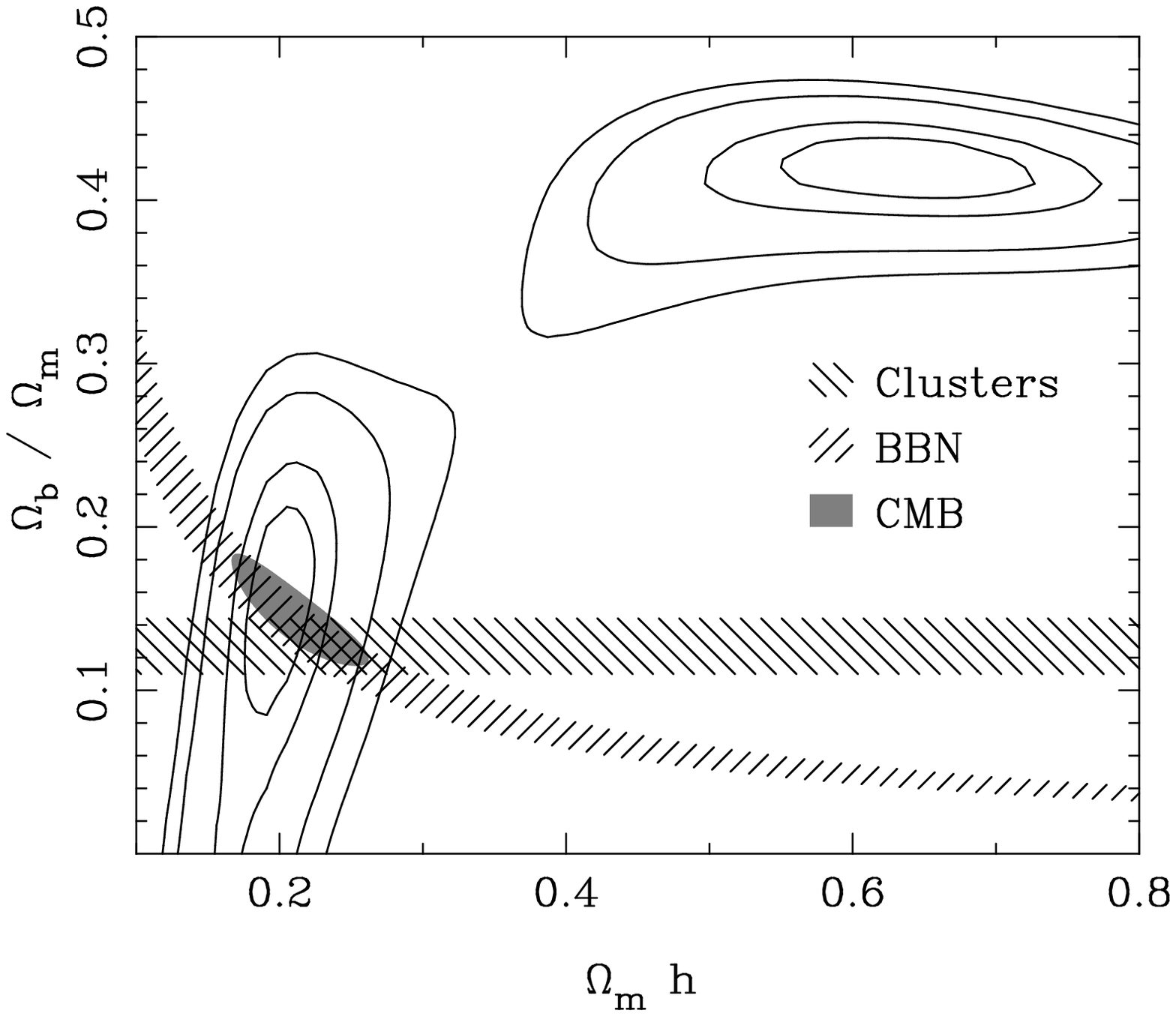}{0.6}
\caption{Likelihood contours for the best-fit linear power spectrum
  over the region $0.02<k<0.15$. The normalization is a free parameter
  to account for the unknown large scale biasing. Contours are plotted
  at the usual positions for one-parameter confidence of 68\%, and
  two-parameter confidence of 68\%, 95\% and 99\% (i.e. $-2\ln({\cal
  L}/{\cal L_{\rm max}}) = 1, 2.3, 6.0, 9.2$). We have marginalized
  over the missing free parameters ($h$ and the power spectrum
  amplitude) by integrating under the Likelihood surface.
  A prior on $h$ of $h=0.7\pm 10\%$ was assumed. 
  This result is compared to estimates from x-ray cluster
  analysis (Evrard 1997), big-bang nucleosynthesis (Burles et al.
  2001) and recent CMB results (Netterfield et al. 2001; Pryke et al. 2002). 
  The CMB results assume that
  $\Omega_bh^2$ and $\Omega_{\rm cdm}h^2$ were independently
  determined from the data.}
\end{figure}

Figure~8 shows that there is a degeneracy between
$\Omega_mh$ and the baryonic fraction $\Omega_b/\Omega_m$. However, there
are two local maxima in the likelihood, one with $\Omega_mh \simeq 0.2$
and $\sim 20\%$ baryons, plus a secondary solution $\Omega_mh \simeq 0.6$
and $\sim 40\%$ baryons. The high-density model can be rejected through a variety
of arguments, and the preferred solution is
$$
  \Omega_m h = 0.20 \pm 0.03; \quad\quad \Omega_b/\Omega_m = 0.15 \pm 0.07.
$$
The 2dFGRS data are compared to the best-fit linear power spectra
convolved with the window function in Figure~9. This
shows where the two branches of solutions come from: the low-density
model fits the overall shape of the spectrum with relatively small
`wiggles', while the solution at $\Omega_m h \simeq 0.6$ provides a
better fit to the bump at $k\simeq 0.065\hompc$, but fits the overall
shape less well.

\begin{figure}[ht]
\plotter{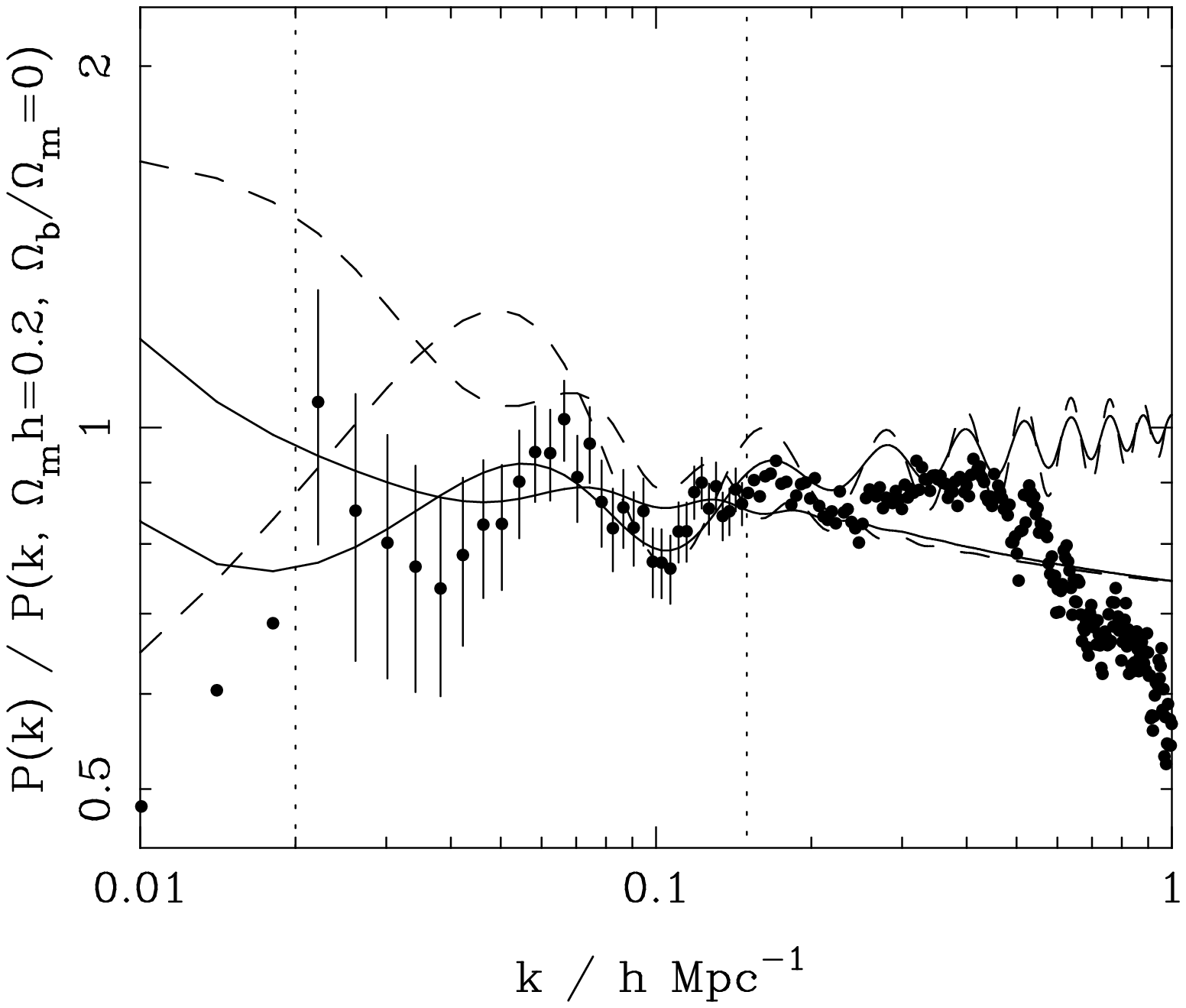}{0.6}
\caption{The 2dFGRS data compared with the two preferred models from
  the Maximum Likelihood fits convolved with the window function
  (solid lines). Error bars show the diagonal elements of the
  covariance matrix, for the fitted data that lie between the dotted
  vertical lines. The unconvolved models are also shown (dashed
  lines). The $\Omega_m h \simeq 0.6$, $\Omega_b/\Omega_m=0.42$,
  $h=0.7$ model has the higher bump at $k\simeq 0.05\hompc$. The
  smoother $\Omega_m h \simeq 0.20$, $\Omega_b/\Omega_m=0.15$, $h=0.7$
  model is a better fit to the data because of the overall shape.}
\end{figure}

It is interesting to compare these conclusions with other
constraints. These are shown on Figure~8, assuming 
$h=0.7\pm 10\%$.
Latest estimates of the Deuterium to Hydrogen ratio in QSO spectra
combined with big-bang nucleosynthesis theory predict $\Omega_bh^2 =
0.020\pm 0.001$ (Burles et al. 2001), which translates to the
shown locus of $f_{\japsub B}$ vs $\Omega_m h$. X-ray
cluster analysis predicts a baryon fraction
$\Omega_b/\Omega_m=0.127\pm0.017$ (Evrard 1997) which is within
$1\sigma$ of our value. These loci intersect very close
to our preferred model. Moreover, 
these results are in good agreement with independent
estimates of the total density and baryon content from
the latest data on CMB anisotropies
(Netterfield et al. 2001; Pryke et al. 2002). For scalar-only
models, these favour CDM and baryon physical densities of
$\Omega_c h^2 = 0.15 \pm 0.03$ and
$\Omega_b h^2 = 0.0215 \pm 0.0025$.
If we take $h=0.7\pm 10\%$,
this gives
$$
  \Omega_m h=0.21\pm 0.05;\quad\quad\Omega_b/\Omega_m=0.14\pm 0.03,
$$
in remarkably good agreement with the estimate from the 2dFGRS
$$
  \Omega_m h = 0.20 \pm 0.03; \quad\quad \Omega_b/\Omega_m = 0.15 \pm 0.07.
$$

Perhaps the main point to emphasise here is that the 2dFGRS results are not
greatly sensitive to the assumed tilt of the primordial spectrum. We
have used CMB results to motivate the choice of $n=1$, 
as discussed below, but it is
clear that very substantial tilts are required to alter the
conclusions significantly: $n\simeq 0.8$ would be required to turn
zero baryons into the preferred model.
The dependence on tilt emphasises that the baryon signal
comes in good part from the overall shape of the spectrum. Although
the eye is struck by a single sharp `spike' at $k\simeq 0.065\hompc$,
the correlated nature of the errors in the $P(k)$ estimate means that
such features tend not to be significant in isolation. We note that
the convolving effects of the window would require a very substantial
spike in the true power in order to match our data exactly. This is
not possible within the compass of conventional models, and the
conservative conclusion is that the apparent spike is probably
enhanced by correlated noise.  A proper statistical treatment is
essential in such cases.

\sec{Combination with the CMB and cosmological parameters}

The 2dFGRS power spectrum contains important information about the
key parameters of the cosmological model, but we have seen that additional
assumptions are needed, in particular the values of $n$ and $h$.
Observations of CMB
anisotropies can in principle measure 
most of the cosmological parameters, and
combination with the 2dFGRS can lift most of the degeneracies inherent
in the CMB-only analysis. It is therefore of interest to see
what emerges from a joint analysis.

These issues are discussed in Efstathiou et al. (2002),
and in Efstathiou's presentation at this meeting.
The CMB data alone contain two important degeneracies:
the `geometrical' and `tensor' degeneracies.
In the former case, one can evade the
commonly-stated CMB conclusion that the
universe is flat, by adjusting both $\Lambda$ and $h$ to
extreme values. In the latter case, a model with a large
tensor component can be made to resemble a zero-tensor model
with large blue tilt ($n>1$) and high baryon content.
Efstathiou et al. (2002) show that adding the 2dFGRS
data removes the first degeneracy, but not the second.
This is reasonable: if we take the view that the CMB determines
the physical density $\Omega_m h^2$, then a measurement of
$\Omega_m h$ from 2dFGRS gives both
$\Omega_m$ and $h$ separately in principle,
removing one of the degrees of freedom on which the geometrical
degeneracy depends. On the other hand, the 2dFGRS alone constrains
the baryon content weakly, so this does not remove the scope
for the tensor degeneracy.

On the basis of this analysis, we can therefore be confident
that the universe is very nearly flat, so it is defensible to assume
hereafter that this is exactly true. The importance of tensors
will of course be one of the key questions for cosmology over the
next several years, but it is interesting to consider the limit
in which these are negligible. In this case, the standard model
for structure formation contains a vector of only 6 parameters:
$$
{\bf p} = (n_s, \Omega_m, \Omega_b, h, Q, \tau).
$$
Of these, the optical depth to last scattering, $\tau$, is almost entirely
degenerate with the normalization, $Q$ -- and indeed with
the bias parameter; we discuss this below.
The remaining four parameters are pinned down very precisely,
as shown in Figure~10. 
Using the latest CMB data, including
COBE (Wright et al. 1996),
Boomerang (Netterfield et al. 2001, de Bernardis et al. 2002), 
Maxima (Lee et al. 2001; Stomper et al. 2001) 
and DASI (Halverson et al. 2002; Pryke et al. 2002), plus the
2dFRGS power spectrum,
we obtain
$$
(n_s, \Omega_c h^2, \Omega_b h^2, h) =
(1.00\pm 0.05, 0.114\pm 0.009, 0.022\pm0.002, 0.68 \pm 0.05),
$$
or an overall density parameter of $\Omega_m=0.29 \pm 0.05$.

It is remarkable how well these figures agree with completely
independent determinations: $h=0.72\pm 0.08$ from the HST key project
(Mould et al. 2000; Freedman et al. 2001);
$\Omega_b h^2 =0.020 \pm 0.001$ (Burles et al. 2001).
This gives confidence that the tensor component must
indeed be sub-dominant.

\begin{figure}[ht]
\plotter{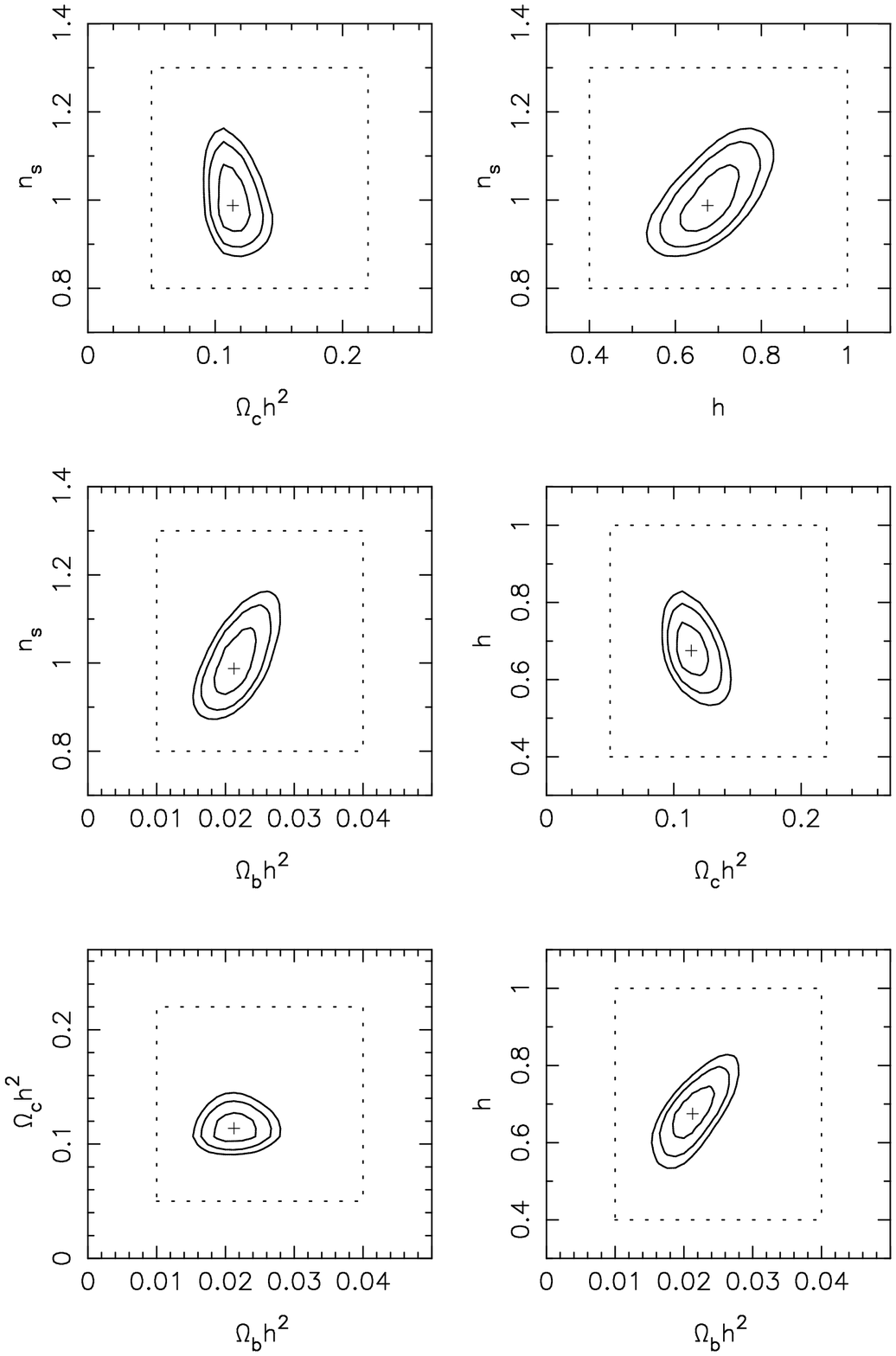}{0.63}
\caption{Likelihood contours for a fit to current CMB data plus the
2dFGRS power spectrum, assuming flat scalar-dominated models.
With these assumptions, there are no significant parameter
degeneracies, and the optimal model is well specified.
Impressively, the required values for the Hubble parameter
and baryon density are in very good agreement with direct
estimates, although no prior information on these parameters was included.}
\end{figure}

\sec{The degree of bias}

\ssec{Bispectrum}

Another key issue which we can address with these results is the extent to which the
galaxy distribution is biased.
As far as the large-scale results on the power spectrum are
concerned,
there is a degeneracy between the
unknown amplitude of the matter power spectrum $P(k)$ and the degree
of bias, $b$, defined such that the galaxy power spectrum is $P_g(k)
\equiv b^2 P(k)$.  
At later times (or on smaller scales),
this degeneracy is lifted by nonlinear effects.  One feature
of nonlinear gravitational evolution is that the overdensity field
becomes progressively more skewed towards high density (assuming
Gaussian initial fluctuations). One could
hope to exploit this
gravitational skewness, but skewness could equally well arise from
biasing, e.g. from a galaxy formation efficiency that increased at
dense points in the mass field.  It is nevertheless possible to
distinguish these two effects by considering the {\it shapes\/} of
isodensity regions.  If the field is unbiased, then the shapes of
isodensity contours become flattened, as gravitational instability
accelerates collapse along the short axis of structures, leading to
sheet-like and filamentary structures.

\begin{figure}[ht]
\plotter{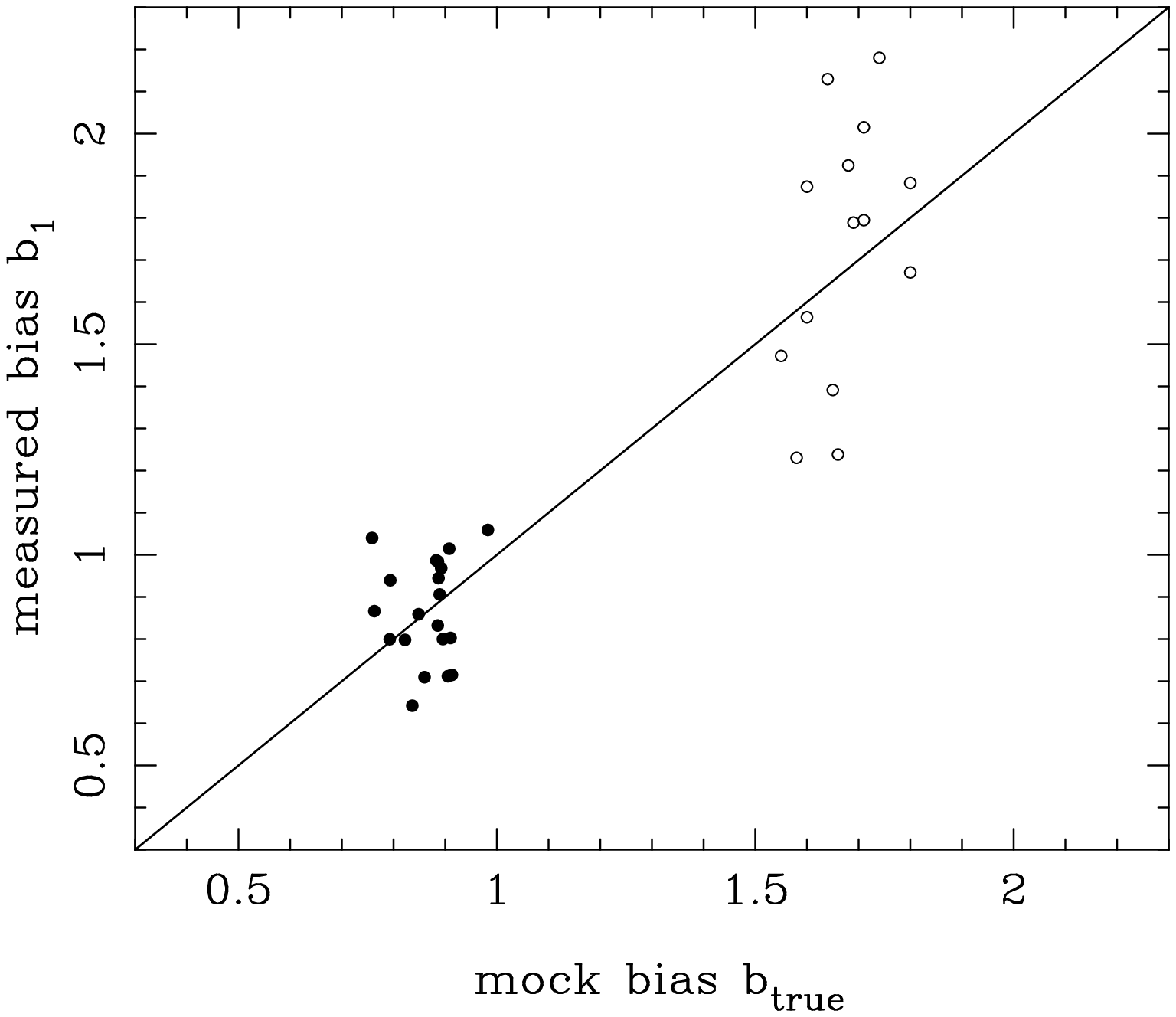}{0.6}
\caption{$b_1$ recovered with the bispectrum method versus the
underlying (true) $b_1\equiv (P_g/P_m)^{1/2}$ for 16 mock SGP simulations
for the $\tau$CDM (open circles) and $\Lambda$CDM (solid circles) models.  
Note that the 2dFGRS has data in the NGP and
SGP, reducing the error bar compared to these mock catalogues.
Even so, it is clear that the bispectrum method can discriminate between
strongly biased $\Omega=1$ models and relatively unbiased models such as $\Lambda$CDM.}
\end{figure}

In Verde et al. (2001), this effect is studied using the
bispectrum $B$, which is related to the three-point correlation
function in Fourier space.  For the mass, this is defined by
$$
\langle \delta_{\bk_1}\delta_{\bk_2}\delta_{\bk_3}\rangle \equiv
(2\pi)^3 B(\bk_1,\bk_2,\bk_3) \delta^{\rm D}(\bk_1+\bk_2+\bk_3),
$$
where $\delta_{\bk} \equiv \int d^3\bx\,
\delta(\bx)\exp(-i\bk\cdot\bx)$ is the Fourier transform of the
mass overdensity and 
$\delta^{\rm D}$ is the Dirac delta function; this shows that
the bispectrum can be non-zero only if the $\bk$-vectors close to
form a triangle.

For Gaussian fluctuations, the bispectrum is zero by
symmetry, but it becomes non-zero under gravitational
evolution, in the same way as a non-zero skewness develops.
In lowest-order perturbation theory, 
Fourier coefficients develop a nonlinear component which is
proportional to $\delta^2$, so the leading order term in the
bispectrum grows like $\delta^4$.  Since the 2dFGRS is not
a survey of mass density, to interpret the bispectrum
measured from the survey we must make some assumption about the
distribution of mass relative to the distribution of galaxies.
The simplest assumption is a bias that is a causal (but
nonlinear) function of the mass density:
$$
\delta_g = b_0 + b_1\delta + b_2 \delta^2/2 + \dots
$$
For consistency with perturbation theory, we must
keep the quadratic bias term involving $b_2$; this
induces a calculable constant offset $b_0$, since
$\langle\delta_g\rangle \equiv 0$. This is a fairly
general formulation, although it ignores stochastic
effects (Dekel \& Lahav 1999), and it is a form
of local bias (modulo the smoothing scale used
to define the quasilinear density field).

For practical estimation of the bispectrum,
it is necessary to decide which triangles
to consider. In practice,  
we use two sets of
triangles of different configurations: one set with one wavevector
twice the length of another, and another set with two wavevectors of
common length.
With a constraint of $k<0.35 \hompc$, this
yields 80 million triangles.
Results for mock data are shown in Figure~11, demonstrating
that highly-biased models can be distinguished from
nearly unbiased models. The mock samples also provide a
direct estimate of the errors on the bias parameters,
yielding the 2dFGRS results:
$$
\eqalign{
b_1 &= 1.04 \pm 0.11 \cr
b_2 &= -0.05\pm 0.08 \cr
}
$$
Remarkably, the simplest possible model works: there 
need be no segregation of mass and light on the largest
scales. On smaller scales, we know that the nonlinear
mass power spectrum has a shape that almost inevitably differs from 
that of the light (e.g. Seljak 2000; Peacock \& Smith 2000),
so light cannot trace mass exactly. Nevertheless, the
degree of large-scale bias is small.

Again, this result is obtained using optimal weighting,
so the value of $b_1=1.04\pm0.11$ refers to $L\simeq 1.9 L^*$
galaxies, as for the $\beta$ analysis. We can therefore
combine these figures to estimate $\Omega_m$, obtaining
$\Omega_m=0.27 \pm 0.06$. Strictly, this refers to $\Omega_m$
at the effective redshift of 0.17; for a flat model, evolution
to $z=0$ is significant, yielding $\Omega_m=0.19 \pm 0.04$ --
a result that is entirely internal to the 2dFGRS. 
This is smaller than the figure of $0.29 \pm 0.05$
from the 2dFGRS+CMB analysis, but the two are consistent;
the formal average
of these two results is $\Omega_m=0.23 \pm 0.03$.

\ssec{Fluctuation amplitude from the CMB}

An alternative and completely independent approach
is to infer the degree of bias directly by using
the amplitude of mass fluctuations inferred from
the CMB. This analysis was performed by Lahav et al. (2001).

\begin{figure}[ht]
\plotter{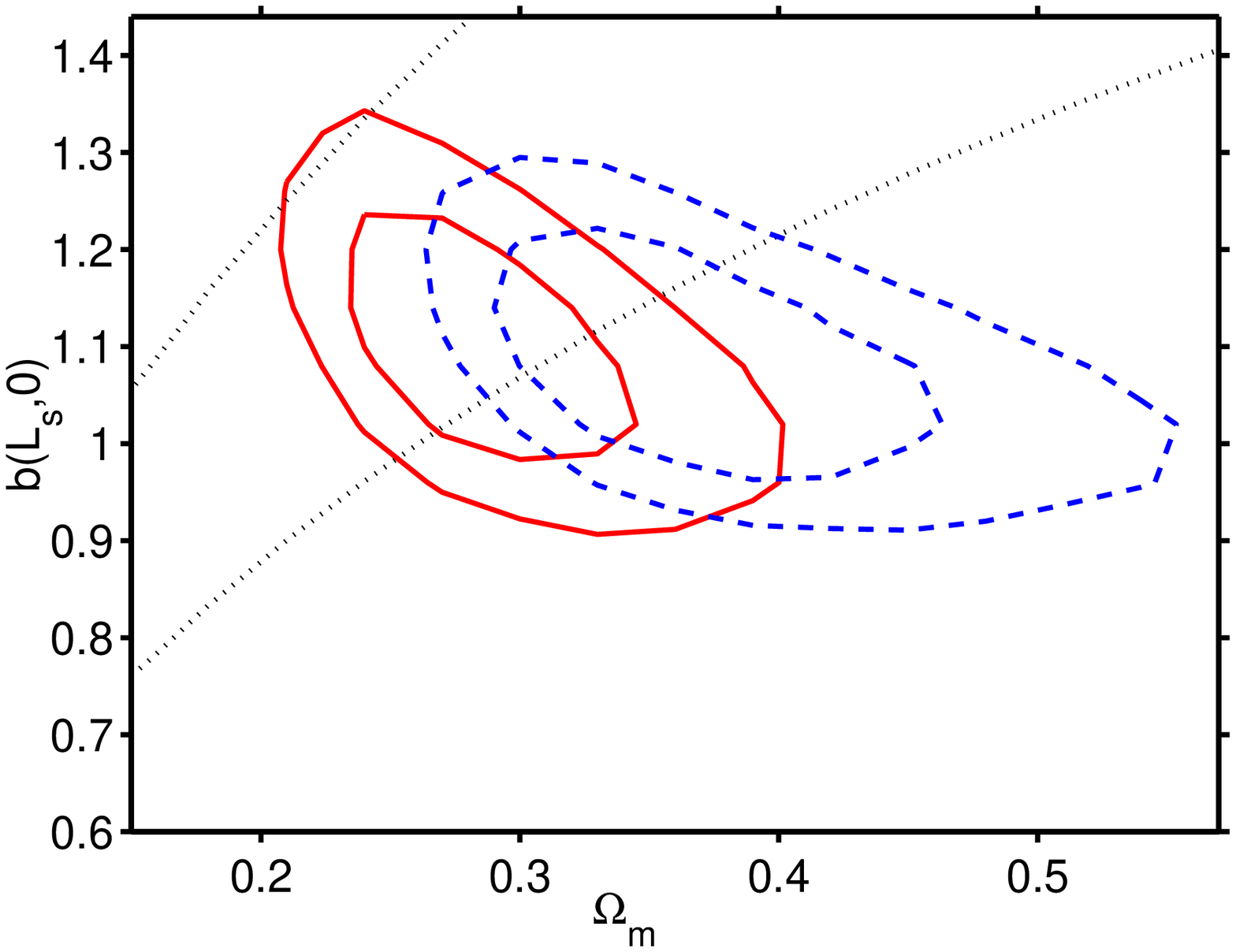}{0.6}
\caption{The result of a joint likelihood fit to 2dFGRS+CMB.
The marginalization (without any external  priors) 
is over $h, \Omega_b h^2$ and 
$\sigma_{8{\rm m}}$. Other parameters are held fixed 
($n=1, \tau = 0$); results for $n=0.9$ are shown as dashed lines.
The contours are for (two-parameter) 68 per cent and 95 per cent confidence intervals.
The  dotted  lines represent a 1-$\sigma$ envelope for
$\beta(L_s,0)$,
based on  $\beta(L_s,z_s) = 0.43 \pm 0.07$ from Peacock et al. (2001)
and the CGC model.
}
\end{figure}

Given assumed values for the cosmological parameters, the
present-day linear normalization of the mass spectrum (e.g. $\sigma_8$)
can be inferred.
It is convenient to define a corresponding measure
for the galaxies, $\sigma_{8{\rm g}}$, such that we can express the bias parameter
as
$$
b = {\sigma_{8{\rm g}} \over \sigma_{8{\rm m}} }.
$$
In practice, we define $\sigma_{8{\rm g}}$ to be the value  required
to fit a CDM model to the power-spectrum data on linear scales ($0.02<k<0.15 \hompc$).
A final necessary complication of the notation is that we need to distinguish
between the apparent values 
of $\sigma_{8{\rm g}}$ as measured in redshift space
($\smash{\sigma_{8{\rm g}}^S}$) and the real-space value that would be measured in the
absence of redshift-space distortions ($\smash{\sigma_{8{\rm g}}^R}$). It is the latter
value that is required in order to estimate the bias.

A model grid covering 
the range
$0.1 < \Omega_m h < 0.3 $, $0.0 < \Omega_b/\Omega_m < 0.4 $, 
$0.4 < h < 0.9$  and $0.75 < \smash{\sigma^S_{8{\rm g}}} < 1.14$
was considered. The primordial index was assumed to be
$n=1$ initially, and the dependence on $n$ studied separately.
For fixed 
`concordance model' parameters
$n=1$, $k=0$, $\Omega_m = 0.3$, $\Omega_b h^2 = 0.02$
and a Hubble constant $h=0.70$, 
we find that the amplitude of 2dFGRS galaxies
in redshift space is $\smash{\sigma_{8{\rm g}}^S} (L_s,z_s) = 0.94$.
Correcting for redshift-space distortions as detailed above
reduces this to 0.86 in real space. Applying a correction for
the mean luminosity of $1.9L^*$ using the recipe of Norberg et al. (2001a),
we obtain an estimate of $\smash{\sigma_{8{\rm g}}^R} (L^*,z_s) = 0.76$,
with a negligibly small random error.
In order to obtain present-day bias figures, we need to know
the evolution of galaxy clustering to $z=0$. Existing data on
clustering evolution reveals very slow changes: higher bias at early
times largely cancels the evolution of the dark matter.
We therefore assume no evolution in $\smash{\sigma_{8{\rm g}}}$:
the CGC (constant galaxy clustering) model.
This has the consequence that $b(z_s) = 1.09\, b(z=0)$.

One can now compare the galaxy figure with $\sigma_8$ for mass,
deduced by fitting to the latest  CMB data, as discussed above.
In Figure~12, we show joint confidence intervals on
the inferred linear bias parameter and $\Omega_m$.
Marginalising over $\Omega_m$ gives an estimate for the linear 
bias parameter. This applies only for $n=1$ and negligible
optical depth to last scattering, but it is simple to obtain
an empirical correction for these effects:
$$
b(L_s, z=0) = (1.10\pm0.08)\, \exp[-\tau + 0.5(n-1)].
$$
Thus, large $\tau$ requires antibias, as does a red
tilt with $n<1$. From earlier results, the tilt effects
are probably negligible. For BBN baryons plus a reionization
redshift of 10, $\tau \simeq 0.07$. Very much larger values
can probably be ruled out, as the bias inferred by this
method would then disagree with the bispectrum calculation.
Formally, taking $n=1$, we obtain $\tau < 0.4$ at 95\% confidence.

\sec{Conclusions}

The 2dFGRS is the first 3D survey of the local universe
to achieve 100,000 redshifts, almost an order of
magnitude improvement on previous work.
The final database should yield definitive results on a
number of key issues relating to galaxy clustering.
For details of the current status of the 2dFGRS, see 
{\tt http://www.mso.anu.edu.au/2dFGRS}.
In particular, this site gives details of the 2dFGRS
public release policy, in which 
approximately the first half of the survey data were made available
in June 2001, with the complete survey database to be made
public by mid-2003. The key results of the survey to date
may be summarized as follows:

\japitem{(1)}The galaxy luminosity function has been measured
precisely as a function of spectral type (Folkes et al. 1999;
Madgwick et al. 2001).

\japitem{(2)}The amplitude of galaxy clustering has been shown
to depend on luminosity (Norberg et al. 2001a). The relative
bias is $b/b^* = 0.85 + 0.15\, (L/L^*)$.

\japitem{(3)}The redshift-space correlation function has been
measured out to $30\mpcoh$. Redshift-space distortions imply
$\beta \equiv \Omega_m^{0.6}/b = 0.43 \pm 0.06$, for
galaxies with $L\simeq 1.9 L^*$.

\japitem{(4)}The galaxy power
spectrum has been measured to high accuracy (10--15\% rms) over about a
decade in scale at $k<0.15\hompc$. The results are very well matched by
an $n=1$ CDM model with $\Omega_mh=0.2$ and 15\% baryons.

\japitem{(5)}Combining the power spectrum results with current CMB data,
very tight constraints are obtained on cosmological parameters. 
For a scalar-dominated flat model, we obtain $\Omega_m=0.29\pm 0.05$,
$\Omega_b h^2 =0.022\pm0.002$
and $h=0.68 \pm 0.04$, independent of external data.

\japitem{(6)}Results from the CMB comparison 
imply a large-scale bias parameter consistent with unity.
This conclusion is also reached in a completely independent way
via the bispectrum analysis of Verde et al. (2001).

\enditem
Overall, these results provide precise support for a cosmological
model that is flat, with $(\Omega_b,\Omega_c,\Omega_v)
\simeq (0.04,0.25,0.71)$, to a tolerance of 10\% in each figure.
Although the $\Lambda$CDM model has been claimed to have problems in matching
galaxy-scale observations, it clearly works extremely well on large scales, and
any proposed replacement for CDM will have to maintain this agreement.
So far, there has been no need to invoke either tilt of
the scalar spectrum, or a tensor component in the CMB.
If this situation is to change, the most likely route will be
via new CMB data,  combined with the key complementary information
that the large-scale structure in the 2dFGRS can provide.

\section*{Acknowledgements}

The 2dF Galaxy Redshift Survey
was made possible by the dedicated efforts of the staff
of the Anglo-Australian Observatory, both in creating the 2dF
instrument, and in supporting it on the telescope.
The 2dFGRS Team thank additional
collaborators whose work is included above:
Sarah Bridle, Alan Heavens, Licia Verde \& Sabino Matarrese.

% References

\section*{References}

\japref Ballinger W.E., Peacock J.A., Heavens A.F., 1996, MNRAS, 282, 877
\japref Baugh C.M., Efstathiou G., 1993, MNRAS, 265, 145
\japref Baugh C.M., Efstathiou G., 1994, MNRAS, 267, 323
\japref Benoist C., Maurogordato S., da Costa L.N., Cappi A., Schaeffer R., 1996, ApJ, 472, 452
\japref Benson, A.J., Frenk, C.S., Baugh, C.M., Cole, S., Lacey, C.G., 2001, MNRAS, 327, 1041
\japref Burles S., Nollett K.M., Turner M.S., 2001, ApJ, 552, L1
\japref Cole S., Kaiser N., 1989, MNRAS, 237, 1127
\japref Colless M. et al., 2001, MNRAS, 328, 1039
\japref Davis M., Geller M.J., 1976, ApJ, 208, 13
\japref Davis M., Peebles, P.J.E., 1983, ApJ, 267, 465
\japref de Bernardis P. et al., 2002, ApJ, 564, 559
\japref Dekel A., Lahav O., 1999, ApJ, 520, 24
\japref Efstathiou G. et al., 2002, MNRAS, 330, L29
\japref Evrard A., 1997, MNRAS, 292, 289
\japref Folkes S.J. et al., 1999, MNRAS, 308, 459
\japref Feldman H.A., Kaiser N., Peacock J.A., 1994, ApJ, 426, 23
\japref Freedman W.L. et al., 2001, ApJ, 553, 47 
\japref Halverson N.W. et al., 2002, ApJ, 568, 38
\japref Hamilton A.J.S., 1992, ApJ, 385, L5
\japref Hamilton A.J.S., Tegmark M., Padmanabhan N., 2000, MNRAS, 317, L23
\japref Kaiser N., 1987, MNRAS, 227, 1
\japref Lahav O. et al., 2001, astro-ph/0112162
\japref Lee A.T. et al.,  2001, ApJ, 561, L1
\japref Lewis I.J. et al., 2002, MNRAS, in press, astro-ph/0202175
\japref Loveday J., Maddox S.J., Efstathiou G., Peterson B.A., 1995, ApJ, 442, 457
\japref Maddox S.J., Efstathiou G., Sutherland W.J., Loveday J., 1990a, MNRAS, 242, 43{\sc p}
\japref Maddox S.J., Sutherland W.J., Efstathiou G., Loveday J., 1990b, MNRAS, 243, 692
\japref Maddox S.J., Efstathiou G., Sutherland W.J., 1990c, MNRAS, 246, 433
\japref Madgwick D. et al., 2001, astro-ph/0107197
\japref Meiksin A.A., White M., Peacock J.A., 1999, MNRAS, 304, 851
\japref Mo H.J., White S.D.M., 1996, MNRAS, 282, 347
\japref Mould J.R. et al., 2000, ApJ, 529, 786
\japref Netterfield C.B. et al., 2001, astro-ph/0104460
\japref Norberg P. et al., 2001a, MNRAS, 328, 64
\japref Norberg P. et al., 2001b, astro-ph/0112043
\japref Outram  P.J., Hoyle F., Shanks T., 2001, MNRAS, 321, 497
\japref Peacock J.A. et al., 2001, Nature, 410, 169
\japref Peacock J.A., Smith R.E., 2000, MNRAS, 318, 1144
\japref Percival W.J. et al., 2001, MNRAS, 327, 1297
\japref Pryke C. et al., 2002, ApJ, 568, 46
\japref Schlegel D.J., Finkbeiner D.P., Davis M., 1998, ApJ, 500, 525
\japref Seljak U., 2000, MNRAS, 318, 203
\japref Stompor R. et al., 2001, ApJ, 561, L7
\japref Taylor A.N., Ballinger W.E., Heavens A.F., Tadros H., 2001, MNRAS, 327, 689
\japref Verde L. et al., 2001, astro-ph/0112161
\japref Wright E.L., Bennett C.L., Gorski K., Hinshaw G., Smoot G.F., 1996, ApJ, 464, L21 

} % end of domain of JAP macros

\end{document}